\documentclass[runningheads]{llncs}
\usepackage[utf8]{inputenc}
\usepackage{graphicx}
\usepackage{booktabs}
\usepackage{amsmath}
\usepackage{amssymb}
\usepackage{cite}
\usepackage{url}
\usepackage{listings}
\let\subparagraph\paragraph
\usepackage{titlesec}
\let\subparagraph\undefined 
\usepackage[width=122mm,left=12mm,paperwidth=146mm,height=193mm,top=12mm,%
paperheight=217mm]{geometry}

\usepackage{enumitem}
\setlist{nosep}
\usepackage{titlesec}
\titlespacing*{\section}
  {0pt}{1.2ex plus .2ex}{0.8ex}
\titlespacing*{\subsection}
  {0pt}{1ex plus .2ex}{0.6ex}
  \usepackage{setspace}
\begin{document}
 \small

\title{From Verification to Herding:\\
Exploiting Software's Sparsity of Influence}

\author{Tim Menzies \and Kishan Kumar Ganguly}
\institute{Department of Computer Science, North Carolina State University\\
\email{timm@ieee.org, kgangul@ncsu.edu}}

\maketitle

\begin{abstract}
Software verification is now costly, taking over half the project effort while failing on modern complex systems. We hence propose a shift from \textit{verification} and modeling to \textit{herding}: treating testing as a model-free search task that steers systems toward target goals. This exploits the "Sparsity of Influence" ---the fact that, often, large   software state spaces are ruled by just a few variables,  We introduce \textit{EZR} (Efficient Zero-knowledge Ranker), a stochastic learner that finds these controllers directly. Across dozens of tasks, EZR achieved 90\% of peak results with only 32 samples, replacing heavy solvers with light sampling.

\end{abstract}

\section{Introduction}
Testing remains the dominant economic bottleneck in software production. 
Despite advances in automation, verification and validation (V\&V) consume up 
to 60\% of total development effort \cite{Long24}. This cost is escalating as 
systems transition from deterministic logic to stochastic behaviors driven by 
AI components, concurrency, and distributed data. In these environments, the 
traditional goal of "verification"---proving the absence of errors across all 
states---is computationally intractable.

A common response has largely been to address this complex problem with yet more    complexity. 
Techniques like symbolic execution, model checking, and fuzzing attempt to 
explore the full combinatorial explosion of the state space ($2^{|A|}$). When 
these fail, researchers turn to higher-order logic systems, such as Answer Set 
Programming (ASP) or Probabilistic Programming (PP), to reason about uncertainty.

We argue that these approaches miss a fundamental physical property of 
software: \textbf{Sparsity of Influence}. While the theoretical state space of 
a program is vast, the effective control space is surprisingly small. 
Empirical evidence across decades of software engineering research suggests 
that program behavior is rarely determined by complex interactions of hundreds 
of variables. Inste ad, it is governed by a "master key" subset of variables 
(often $|A'| \le 10$).

This paper proposes \textbf{Herding}, a change to the nature of the testing 
process. Instead of verifying correctness or building complex  models, 
Herding uses lightweight sampling to identify these "master keys" and steer 
the system toward a "Heaven" state (e.g., zero defects, low latency).

Our contributions are:
\begin{itemize}
    \item \textbf{The Anti-Modeling Argument:} We critique the "Model-First" 
    approach (ASP/PP), since  direct data sampling can be more cost-
    effective than building new kinds of models (Section 2).
        \item \textbf{The Generalization
        of ``Testing'' }  Section 3 makes a theoretical argument that the prpcess we call ``testing'' cab be ge generatized to cover a large space of tasks. 
    \item \textbf{The Sparsity Synthesis:} We compile extensive evidence 
    showing that software sparsity is a general result, present in many domains  (Section 4).
    \item \textbf{The EZR Recipe:} We present a 
    recipe for EZR (Efficient Zero-knowledge Ranker), an algorithm that 
    exploits sparsity to optimize software with minimal compute (Section 5).
    \item \textbf{Empirical Evidence:} We show that on 63 tasks, 32 samples are 
    sufficient to reach 90\% of  optimality (Section 6).
\end{itemize}

\section{The Modeling Trap }
Before detailing our approach, we must address the prevailing "high-road" 
solutions to software uncertainty: Answer Set Programming (ASP) and 
Probabilistic Programming (PP). Both communities offer sophisticated tools 
to handle non-determinism. However, both fall into the "Modeling Trap."

\subsection{The Cost of $T$ (The Theory)}
Both ASP and PP rely on the existence of a robust model, $T$.
\begin{enumerate}
    \item \textbf{Answer Set Programming (ASP):} ASP solves search problems 
    by reducing them to stable models of logic programs \cite{Brewka2011}. While 
    powerful for well-defined combinatorial problems (e.g., graph coloring, 
    scheduling), applying ASP to testing requires encoding the system's 
    behavior into logical rules. 
    \item \textbf{Probabilistic Programming (PP):} PP frameworks like Stan 
    \cite{carpenter2017stan} or Pyro \cite{bingham2019pyro} allow developers to 
    specify probabilistic models and perform inference (usually MCMC). 
    However, PP requires a model  $T$ that defines priors and 
    relationships.
\end{enumerate}

The drawback in both approaches is the assumption that $T$ is cheaper to build 
than the software itself. In modern software engineering (SE), the system 
\textit{is} the model. Recreating a logical or probabilistic shadow of a 
million-line distributed system is a "boil the ocean" strategy. It shifts the 
burden from "testing the code" to "verifying the model of the code."

\subsection{The "Data-First" Alternative}
Our approach posits that we do not need to fully \textit{understand} a model 
chain to \textit{control} it. This is a "black-box" optimization philosophy. 
If the control space is sparse, we can ignore the complex internal mechanics 
($T$) and focus solely on the Input/Output pairs $(X, Y)$.

Instead of asking "Why does this fail?" (which requires a Model), Herding 
asks "Which inputs maximize success?" (which requires only data). By bypassing 
the construction of $T$, we eliminate a   fragile and expensive step in 
the V\&V pipeline. As we show in Section 6, simple rank-based sampling can
outperform more complex reasoning because it goes straight to the empirical 
reality of the system.

\section{An Abductive Model of Testing (and Other Things)}

To formalize this ``data-first'' stance without falling back into
the ``modeling trap,'' we adopt Poole's abductive
framework~\cite{Poole1994}. Unlike the heavy model-building of ASP,
Poole's logic allows us to accept the system-as-is as the theory ($T$)
and reframe the engineering challenge as the optimization of
\textit{assumptions} ($A$).
In turns out that Poole's framework provides
a unifying theory for activity across the  software engineering life cycle, from
requirements engineering (at inception) to verification (at the end). It handles an interesting and challenging case (testing non-deterministic systems) while also offering engineering insight into when many SE tasks, including testing, can become very simple indeed (specifically, when the number of key assumptions is very small).

(In this paper, \textit{assumptions} are an umbrella term that covers design choices, runtime choices coded by a programmer, or environmental factors  outside the program e.g., traffic on the internet, weather.)

In summary, in this framework, all inference becomes an optimization task exploring assumptions in order to reach goals.
In the general case, only a subset of a theory $T$ can be used to
achieve goals $G$ using assumptions $A$ without leading to errors
(denoted $\bot$):

\begin{equation}\label{eq:subs}
\begin{array}{l@{\hspace{1cm}}l}
 T  \subseteq \mathit{theory}       & T \wedge A  \vdash  G        \\
 A \subseteq \mathit{assumptions}  & T \wedge A  \not\vdash \bot \\
 G \subseteq \mathit{goals}        &
\end{array}
\end{equation}

\noindent In plain English:
\begin{itemize}
    \item $ T \wedge A \vdash G$ means testing that ``we can do
    something'' (functionality);
    \item $T \wedge A \not\vdash \bot$ means testing that ``we don't
    mess up'' (validity).
\end{itemize}

Equation~\ref{eq:subs} is a useful model for a variety of tasks. {\em Nondeterminism}, is  the stochastic
selection among these different worlds
 In  {\em design}, some humans choose the assumption, while
 in {\em diagnosis}, the world chooses $A$, and our task is to find the minimal $A$ that covers the symptoms $G$. 

Equation~\ref{eq:subs} also models common SE tasks.
 {\em Testing} is  the search for an assignment $A'$ that triggers a specific behavior (e.g., a crash or a successful transaction).  More specifically, {\em metamorphic testing}  asks the question: what extra knowledge can be imported to assess a system; i.e., it focuses on the clause
$T \wedge A  \not\vdash \bot 
 G \subseteq \mathit{goals}$.
 
But not just testing: Equation~\ref{eq:subs} maps cleanly onto formal models of 
requirements engineering.
The traditional view of
RE~\cite{Zave1997} assumed that Specification ($S$) and Domain Knowledge
($K$) must logically entail Requirements ($R$). However, Jureta et
al.~\cite{Jureta2008} argue in their CORE framework that this binary
view fails to account for \textit{softgoals} (preferred but not
essential goals). They propose RE is an optimization task:
\begin{equation}\label{core}
K, S \vdash_{sat} G \quad \text{and} \quad \text{val}(K, S, P)
\text{ is optimal}
\end{equation}
Note that Poole's  theory $T$ maps to Domain Knowledge $K$,
while CORE splits assumptions $A$ into Specification $S$ and
Preferences $P$. 
Our reading of the CORE
paper is that:
\begin{itemize}
\item
They view  assumptions exploration as a  design-time activity that results in a single system;
\item
While we see assumption exploration as   a continuous process that persists throughout design and runtime and testing.
\end{itemize}
How does all  this let us avoid
the modeling trap?
Note that we do not need to know $T$ in order to find the $A'$ values that control a system
As shown below,  we can treat $T$ as a black box, generating
samples of input assumptions and output results. Then it becomes a data mining task
to  find $A'$ without $T$ by  sampling the
\textit{contrast} between the best and worst system behaviors.

The most important aspect of 
Equation~\ref{eq:subs} is that it tells
us when we can efficiently complete all the above tasks (design, diagnosis, testing, requirements engineering, all in the presence of nondeterminism).
Let the \textit{key assumptions} ($A'$) be the subset of $A$ that
operates independently of other assumptions. ``Worlds of belief'' $W
\subseteq A'$ represent minimal consistent beliefs (alternative
possible executions).   
When there are many subsets, 
Equation~\ref{eq:subs} is slow to compute
(actually, NP-hard~\cite{Bylander1991}). On the other hand, when $|A'|$ is tiny,
Equation~\ref{eq:subs} terminates quickly.

So,  in practice, how big is $|A'|$?
And can we build data miners that find it?
The next section offers evidence that, empirically,  $|A'|$ is very small.
After that, the EZR contrast set learner is discussed.

\section{Evidence for Sparsity}
The viability of "Model-Free" Herding rests entirely on the existence of 
Sparsity. If software were truly a high-dimensional chaotic system where 
every variable interacted with every other variable, we would need massive 
models (ASP/PP) or massive test suites.

Fortunately, software is physically "thin." We present evidence from four 
abstraction layers: Logic, Code, Runtime, and Design.

\subsection{Layer 1: Logic and SAT}
At the lowest level of abstraction, software verification problems can often be
encoded as Satisfiability (SAT) instances. Theoretically, SAT is NP-complete, 
implying exponential worst-case complexity ($2^N$).

 However, in a landmark study, Selman et al. \cite{Selman03} discovered 
"backdoors" in SAT instances derived from real-world software. They found 
that while the variables numbered in the thousands, the logical structure was 
controlled by a "backdoor" set of fewer than a dozen variables.  The time to find
solutions can be dramatically reduced by
setting these backdoors before execution.
As shown in Table~\ref{tab:selman}, this effect is not subtle. In
real-world benchmarks ranging from logistics planning to hardware
verification, the number of backdoor variables is vanishingly small
compared to the total problem size.
\begin{table}[!h]
\centering
\footnotesize
\caption{Backdoor sizes in standard benchmarks (adapted from
~\cite{Selman03}). Note the dramatic reduction from total variables
($N$) to the size of the backdoor set ($|A'|$).}
\label{tab:selman}
\begin{tabular}{l|r|r}
\textbf{Benchmark} & \textbf{Total Vars ($N$)} & \textbf{Backdoor Size
($|A'|$)} \\
\midrule
logistics.d     & 6,750  & 12 \\
3bit-adder\_32  & 8,704  & 53 \\
pipe\_07        & 1,685  & 23 \\
qcp\_15         & 3,375  & 290 \\
map\_v6         & 3,929  & 167
\end{tabular}
\end{table}

This suggests 
that "complexity" in logic may arise
from overlooking a  small core of 
most critical decision points.   Analagous results
have been seen from Holzmann-style state spaces~\cite{pelanek2008properties}.

\subsection{Layer 2: Source Code Structure}
Moving over to source code, we see the same phenomenon. Hindle et al. 
\cite{Hindle12} famously demonstrated that software is "Natural."; i.e.  code is 
written by humans for humans,  humans have limited working memory 
(Miller's Law \cite{miller1956magical}),
so the code people write it actually
surprisingly regular with many repeated patterns.

This cognitive bottleneck forces code to follow strict regularities, adhering 
to Zipfian distributions. Consequently, complexity is not uniformly 
distributed. Ostrand et al. \cite{Ostrand04} and Hamill \cite{Hamill09} 
analyzed large industrial codebases and found that 20\% of the files 
contained 80\% of the defects. Furthermore, within those files, the 
defects were clustered. This "Pareto Principle of Bugs" implies that complete search across
the whole space
 is misguided, while directed sampling (Herding) can be incredibly 
potent if it targets the "natural" clusters.

This structural regularity produces concentration effects that have been observed
from multiple angles in software engineering:
\begin{itemize}
  \item \textbf{Knowledge distribution.} Developers typically work in and understand only
  a small fraction ($\approx 20\%$) of a large codebase, yet systems function coherently \cite{Peters15}.
  \item \textbf{Change distribution.} Fine-grained code changes follow heavy-tailed/power-law
  patterns, suggesting development activity (and therefore risk) is localized rather than uniform
  \cite{Lin15}.
  \item \textbf{Reuse distribution.} A large fraction (up to 52\%) of commits reuse existing code tokens \cite{martinez2014fix}. This empirically indicates that software evolution exhibits structural sparsity.
\end{itemize}

\subsection{Layer 3: Runtime Behavior}
Sparsity persists even during execution. Two areas confirm this: Mutation 
Testing and Fuzzing.
\begin{itemize}
    \item \textbf{Mutation Testing:} Michael and Jones 
    \cite{michael1997uniformity} and Wong \cite{wong1995reducing} found that 
    the vast majority of artificial faults (mutations) injected into code 
    do not propagate to the output. The system absorbs the noise. Conversely, 
    the faults that \textit{do} propagate tend to be catastrophic and easy to 
    trigger. This "Uniformity of Error Propagation" means we do not need to 
    test every path; we only need to test the paths that matter.
    \item \textbf{Big Data Fuzzing:} In the context of Big Data analytics, 
    Zhang et al. \cite{Zhang20BigFuzz} showed that despite processing petabytes 
    of data, the application logic typically exercises fewer than 50 distinct 
    execution paths. That is, even in the input data changes,  the logic paths remain sparse and few in number.
\end{itemize}

\subsection{Layer 4: Requirements and Design}
Finally, at the highest level, requirements engineering reveals that stakeholders 
care about very few "Key" distinctives. 
 We have observed this phenomenon in
large-scale requirements models models:

\begin{itemize}
    \item \textbf{NASA Deep Space:} In requirements models for NASA
    deep-space missions, Feather and Menzies found that optimization
    could converge rapidly by finding a few key decisions that
    controlled the objective function~\cite{Feather02}.

    \item \textbf{Modernization Roadmaps:} In the SEI's software modernization roadmaps \cite{ernst2016}, models were controlled
    by a small set of variables (less than 12\% of decisions). Once these
    were set, the remaining search space collapsed, accelerating
    reasoning by orders of magnitude~\cite{Mathews17}.
\end{itemize}
Menzies et al. \cite{MenziesKeys21} 
summarized years of data mining studies to show that in datasets with hundreds 
of attributes, classifiers could predict outcomes with high accuracy using 
only less than 10 attributes.

In {\bf summary}, the "Sparsity of Influence" is not an accident; it is a 
necessary condition for software engineering. If software were not sparse, 
it would be incomprehensible to the humans who write it. Herding simply 
exploits this human limitation.

\section{Methodology: The EZR Recipe}
We propose \textbf{EZR} (Efficient Zero-knowledge Ranker), a "recipe" for 
exploiting sparsity. EZR is a stochastic contrast set learner.
It assumes that if a system is controlled by a small number of effects, then   a   sample of behavior (ranging from ``good'' to ``bad'') will contain those controllers in the  contrast set of the independent variables values seen in ``good'' and ``bad''. 

EZR differs 
from genetic algorithms (which mutate populations) and gradient descent 
(which requires differentiability). EZR is a \textit{discretized} sampler 
that herds the system into "Heaven."

EZR was inspired by TPE~\cite{Bergstra11}, that divides the known solutions into two good and bad sets.  While TPE models those sets with mutiple kernel functions, EZR uses a simple two class Bayes classifier.

EZR was also motivated by the computational problems of SMAC~\cite{hutter2011sequential}. SMAC (and TPE and EZR) are
sequential model optimization algorithms that use
the model built so far to decide where to sample
next. SMAC's model are  multiple decision trees in a random forest. This forest is rebuilt whenever new data is encountered. EZR, on the other hand, uses Welford's algorithm to incrementally adjust its models without having to reset and rebuild from scratch. This make EZR
orders of magnitude faster than SMAC (for 20 runs over all the data shown below, EZR terminated in minutes while SMAC terminates in days).

\subsection{Formalism: Herding vs. Verification}
Let $S$ be the software system with inputs $X$ and outputs $Y$.
\begin{itemize}
    \item \textbf{Verification:} Find $\forall x \in X: \text{Valid}(S(x))$.
    \item \textbf{Herding:} Find $x' \in X$ such that $Utility(S(x')) > \tau$, 
    where $\tau$ is an acceptability threshold ("Heaven").
\end{itemize}
Herding is a satisficing task \cite{Simon1956};
i.e. it an optimizer that finds near enough good solutions.
We define "Heaven" via a 
multi-objective loss function $D(x)$, the "Distance to Heaven." (see below).

\subsection{The EZR Algorithm}
EZR operates by iteratively building a "Contrast Set"---rules that 
distinguish the best current solutions from the rest.

\begin{enumerate}
    \item \textbf{Setup:} Define Objectives $O = \{o_1, o_2, ...\}$. 
    Ideally, $o_i=0$ (e.g., zero bugs, zero delay).
    
    \item \textbf{Initialization:} 
    Sample $N=4$ random configurations $x$ from the input space.
    
    \item \textbf{Scoring (The Loss Function):}
    Calculate $D(x)$, the normalized Euclidean distance to the ideal point:
    \begin{equation}
        D(x) = \frac{\sqrt{\sum_{i \in O} (o_i(x) - \text{ideal}_i)^2}}{\sqrt{|O|}}
    \end{equation}
    Normalization is critical to handle multi-objective trade-offs (e.g., 
    latency in ms vs. defects in integers).

    \item \textbf{Splitting (The Ranking):}
    Sort the current population by $D(x)$ ascending.
    Split into two groups:
    \begin{itemize}
        \item $BEST$: The top $\sqrt{N}$ samples.
        \item $REST$: The remaining $N - \sqrt{N}$ samples.
    \end{itemize}

    \item \textbf{Discretization (The Modeling Surrogate):}
    For every attribute $attr$ in the input space, discretize the values 
    into ranges (bins) based on the population frequencies. This creates 
    a non-parametric distribution of values.
    
    \item \textbf{Acquisition (The Contrast):}
    We seek the attribute range $r$ that maximizes the probability of being 
    in $BEST$ while minimizing the probability of being in $REST$. We 
    calculate the score for every range:
    \begin{equation}
        Score(r) = P(r | BEST)^2 / (P(r | REST) + \epsilon)
    \end{equation}
    Select the rules (attribute ranges) with the highest scores.  

    \item \textbf{Generation:}
    Generate new samples $X_{new}$ by imposing the selected rules (constraints) 
    and randomly sampling the remaining unconstrained variables. This 
    "Herds" the sampling toward the sparse controller region.

    \item \textbf{Loop:}
    Add $X_{new}$ to the population. Increment $N$. Repeat until $N_{max}$ 
    or $D(x) < \text{threshold}$.
\end{enumerate}

\section{Experiments}
We hypothesize that because of sparsity, 
then number of samples $N_{max}$ required to find optimial solutions\footnote{Here, ``optimal'' means ``reference optimal,''
    which is a term from empirical algorithms
    literature~\cite{McGeoch2012,cohen1995empirical}. The reference
    optimal refers to the best solution observed so far. The
    reference optimal may not be the true optimal. However, in many
    engineering cases, the true optimal may be unknown.} can be extremely small.
We test this on the \textbf{MOOT Repository} \cite{Menzies2025MOOT}.

\begin{table*}[p]
\scriptsize
\centering
\caption{\textbf{The MOOT Repository: Dataset Characteristics.}
Detailed breakdown of the 63 optimization tasks. 
$|A|$: Decision variables (inputs). 
$|O|$: Objective variables (goals). 
$N$: Sample count (rows).
\textit{Opt\%}: Mean performance of EZR at $N=32$ vs reference optimal.}
\label{tab:moot_full}
\setlength{\tabcolsep}{3pt} 
\renewcommand{\arraystretch}{1.2}

\begin{tabular}{@{} p{1.6cm} p{4.8cm} p{1.5cm} r r r r @{}}
\toprule
\textbf{Family} & 
\textbf{Optimization Goal \& Description} & 
\textbf{Ex.} & 
\textbf{$|A|$} & 
\textbf{$|O|$} & 
\textbf{Rows} & 
\textbf{Opt\%} \\
\midrule

\multicolumn{7}{@{}l}{\textit{\textbf{I. System Configuration}}} \\ 
\raggedright Sys.\ Config & 
\raggedright \textbf{Cost/Latency.} Tuning parameters for hardware 
deployments to minimize delay. & 
SS-* & 
3-88 & 2 & 
86k & 
91 \\

\raggedright Perf.\ Tuning & 
\raggedright \textbf{Throughput.} Flag tuning for compilers (LLVM), 
DBs, and video compression. & 
x264 & 
9-35 & 1 & 
166k & 
89 \\

\raggedright Cloud & 
\raggedright \textbf{Efficiency.} Tuning server stacks (Apache, SQL) 
for high-load environments. & 
SQL & 
9-39 & 1 & 
4k & 
92 \\

\midrule
\multicolumn{7}{@{}l}{\textit{\textbf{II. Process \& Planning}}} \\ 
\raggedright Scrum & 
\raggedright \textbf{Feature Selection.} Optimizing delivery value 
vs. effort in agile models. & 
Scrum & 
124 & 3 & 
100k & 
88 \\

\raggedright Risk & 
\raggedright \textbf{Mitigation.} Balancing effort against potential 
defects (NASA93, COCOMO). & 
XOMO & 
27 & 4 & 
10k & 
94 \\

\raggedright Health & 
\raggedright \textbf{Activity.} Forecasting project vitality based 
on PRs and commit velocity. & 
Issues & 
5 & 2 & 
10k & 
90 \\

\raggedright SPL & 
\raggedright \textbf{Product Lines.} Solving constraints for valid 
configurations in feature models. & 
FM-* & 
1000 & 3 & 
10k & 
85 \\

\midrule
\multicolumn{7}{@{}l}{\textit{\textbf{III. Analytics}}} \\ 
\raggedright Finance & 
\raggedright \textbf{Churn.} Predicting customer/user churn to 
optimize retention strategies. & 
Bank & 
77 & 3 & 
20k & 
93 \\

\raggedright Talent & 
\raggedright \textbf{Retention.} Analyzing employee attrition 
patterns to improve stability. & 
HR & 
55 & 2 & 
17k & 
89 \\

\raggedright Safety & 
\raggedright \textbf{Critical.} Medical software predictions 
(COVID19) where precision is key. & 
Hosp & 
64 & 2 & 
25k & 
91 \\

\raggedright Sales & 
\raggedright \textbf{Trends.} Optimizing product positioning based 
on marketing analytics. & 
Mkt & 
31 & 8 & 
2k & 
87 \\

\midrule
\multicolumn{7}{@{}l}{\textit{\textbf{IV. Experimental}}} \\ 
\raggedright Control & 
\raggedright \textbf{RL Tasks.} Standard reinforcement learning 
tasks mapped to static optimization. & 
Pole & 
11 & 4 & 
318 & 
95 \\

\raggedright Testing & 
\raggedright \textbf{Reduction.} Selecting optimal test cases to 
maximize coverage/faults. & 
T120 & 
9 & 1 & 
5k & 
92 \\

\bottomrule
\end{tabular}
\end{table*}
\subsection{Experimental Setup}
\textbf{Data:} As seen in Table~\ref{tab:moot_full},
MOOT contains 63 optimization tasks from real-world SE scenarios, 
including:
\begin{itemize}
    \item \textbf{LLVM:} Tuning compiler flags for binary size/speed.
    \item \textbf{X264:} Video encoding parameter tuning.
    \item \textbf{POM3:} Agile project management simulation.
    \item \textbf{Health:} Predicting software project health.
\end{itemize}

\textbf{Metric:} We use \textbf{Normalized Regret} to compare performance 
across different domains. Let $y_{opt}$ be the known global optimal (or best 
known) value, $y_{av}$ be the average of random sampling, and $y_{ezr}$ 
be the result of our approach, the we 
 report the "\% of Optimality," calculated as $1 - Regret$ (normalized).
\begin{equation}\label{regret}
    Regret = 1 - \frac{y_{ezr} - y_{opt}}{y_{av} - y_{opt}}
\end{equation}
This normalization allows us to compare results from different data sets\footnote{A recent  IEEE TSE article by Chen et al. \cite{ref41} reviewed various multi-objective optimization performance measures like Hypervolume (HV), Spread ($\Delta$), Generational Distance (GD), and Inverted Generational Distance (IGD).
But based on recent results with   PROMISETUNE   (where for MOOT-like problems, most of the good solutions were found in very small region~\cite{Chen26Promise}), Chen has moved away from    metrics that look across a wide space (like HV, spread, GD, IGD)  since most of  valuable solutions occur in a small part of that space.}.


\section{Results}
\begin{figure*}[!t]
  \centering
\includegraphics[width=.7\linewidth]{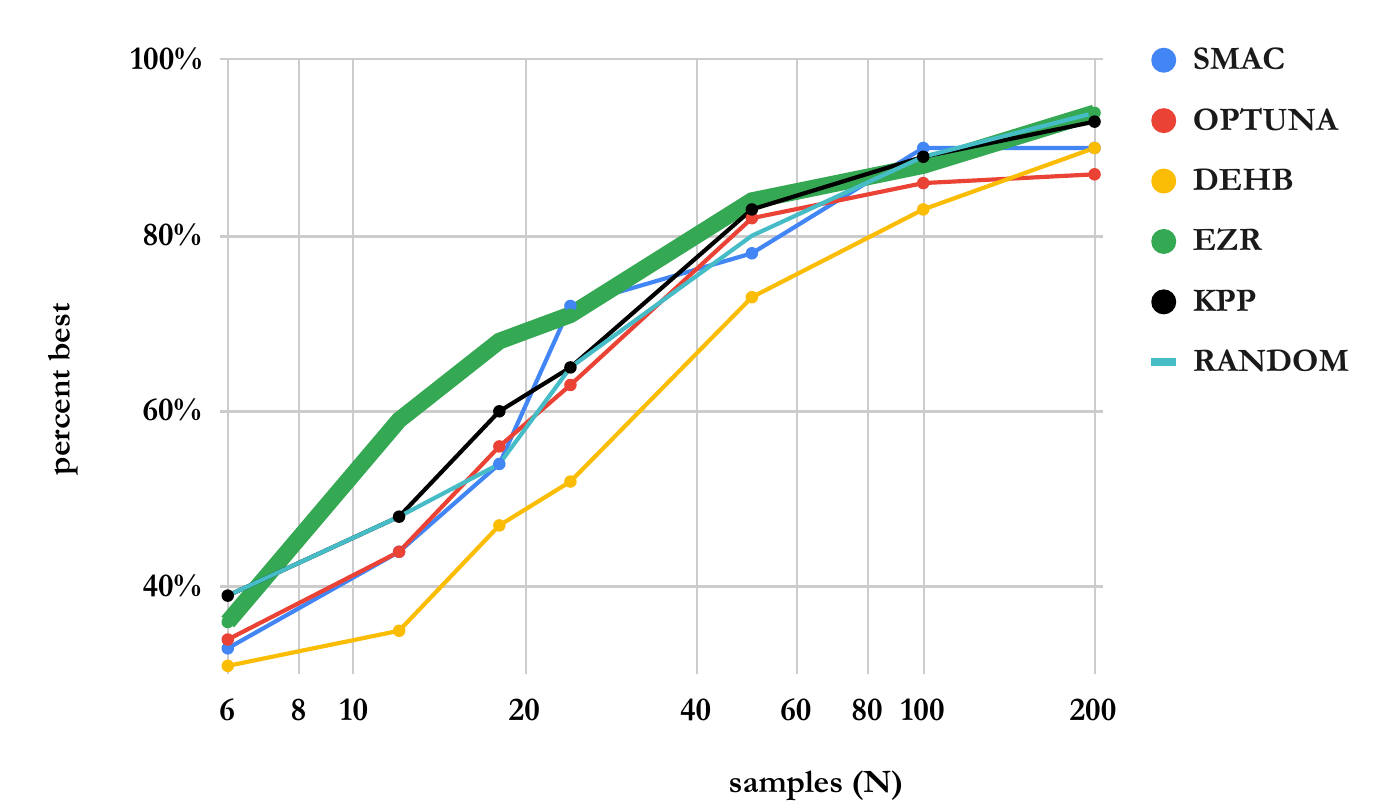}
  \caption{\small EZR versus SMAC and other algorithms. Experiments on the   MOOT data~\cite{Menzies2025MOOT}.  Statistical analysis of 20 runs with different random seeds with scores generated by Equation~\ref{tab:results}.
  Results ranked using nonparametric effect size (Cliff's Delta) and significance tests (KS-test, 95\% confidence). The $y$-axis shows how often one optimizer   was statistically as  good as, or better than, all other algorithms
  (so {\em larger} y=values  are {\em  better}). All plots rise to a high value on the right-hand-side since, given enough samples, the performances of different algorithms mostly tie.  Note that EZR (the green line) performs as well or better than SMAC (and everything else). From~\cite{datalite}.}
  \label{fig:smac}
\end{figure*}

Ganguly and Menzies~\cite{datalite} compared EZR against SMAC and
several other state-of-the-art algorithms on high-dimensional
software configuration tasks. As shown in Figure~\ref{fig:smac}, EZR
outperforms or matches:
\begin{itemize}
    \item \textbf{OPTUNA~\cite{akiba2019optuna}:} A framework
    extending Hyperopt's Tree-structured Parzen Estimators (TPE);
    \item \textbf{DEHB~\cite{awad2021dehb}:} A hybrid of Differential
    Evolution and Hyperband for multi-fidelity optimization;
    \item \textbf{Random:} Simple random selection of $N$ items;
    \item \textbf{KPP~\cite{Arthur2007}:} A variant of random search
    using K-means++ clustering to select distinct centroids.
\end{itemize}

Figure~\ref{fig:smac} reports relative results
of one method ranked against others.
Table \ref{tab:results} presents comments
on the absolute values of those results,  across all 63 
datasets.
\begin{table}[h]
\centering
\caption{Performance of EZR across 63 MOOT tasks. "Optimality" indicates 
proximity to the reference optimal solution (100\%).}
\label{tab:results}
\begin{tabular}{l|c|c}
\textbf{Sampling Budget ($N$)} & \textbf{\% of Optimality}    \\
\midrule
8 samples  & 62\%   \\
16 samples & 80\%  \\
32 samples & 90\%   \\
64 samples & 91\%   \\
128 samples & 92\%   \\
\end{tabular}
\end{table}
The most striking result is the sharp "knee" at 32 samples.
\begin{itemize}
    \item \textbf{0 to 16 Samples:} The algorithm rapidly learns the gross 
    geography of the search space. It eliminates obviously bad regions.
    \item \textbf{16 to 32 Samples:} EZR locks onto the core variables. 
    It identifies the 2--3 variables that most control the objective 
    function.
    \item \textbf{32 to 128 Samples:} Returns diminish rapidly. Doubling the 
    budget from 32 to 64 yields only a 1\% improvement. Quadrupling to 128 
    yields another 1\%.
\end{itemize}

This behavior empirically proves the Sparsity Hypothesis. If the space were 
dense (complex), we would expect a linear or log-linear learning curve 
requiring hundreds of samples. The fact that learning saturates at $N=32$ 
implies that the "information content" regarding the optimal solution is 
fully contained in a very small subset of the data.

\section{Discussion}
\subsection{The "Good Enough" Principle}
Critics might argue that 90\% optimality is not 100\%. For mission critical and safety-critical applications, there is a case for ignoring  sparsity making more samples of the system.
That said:
\begin{itemize}
\item For many engineering domains, the ``near-enough'' results from Table \ref{tab:results} would suffice.
\item Even in mission-critical and safety-critical domains, as   systems grow larger and more complexity, some heuristic method  (like EZR) is required to assess that software.    
\end{itemize}


\subsection{Why Does all this Work?}
Why is software so sparse? We believe it is an artifact of iterative 
software construction. 
Consider a large system with modules $A, B, C$. If Developer 1 understands
Module $A$, they are more likely to teach Developer 2 about $A$ than $B$ or
$C$. This creates self-reinforcing knowledge loops where development activity
(and consequently, bugs and complexity) clusters in small ``corners'' of the
system. This explains the prevalent power laws in software engineering; for
example, Hamill and Goseva-Popstojanova showed that the vast majority of errors
in software are found in a very small region of the codebase~\cite{Hamill09}.

To say that another way,  sparsity of influence may exists   because  human cognitive
    limits restrict complexity. Large Language Models (LLMs) face no such
    biological constraints and is free to generate very different kinds of code. Will AI-generated software evolve into "Alien Code"
    with dense, high-dimensional dependencies ($|A'| \gg 10$) that defy
    herding? Future work must track the dimensionality of AI code to see if we
    need to enforce "Sparsity Constraints" as a safety requirement for
    generative AI.

\section{Conclusion}
The complexity crisis in software testing is partially self-inflicted. By 
treating software as a generic high-dimensional black box, we force ourselves 
to use high-dimensional tools (Verification, ASP, PP) that can scale poorly.

By accepting the existance of \textbf{Sparsity of Influence}, we 
can downgrade the problem difficulty. We do not need to verify everything; 
we only need to control a few key variables.  
The EZR recipe demonstrates that this is not just theory. In the data sets explored here, a budget of 
32 samples can solve optimization problems that seemingly require 
complex solvers.

Our recommendation to the community is simple: before you build a model ($T$), 
try Herding the data ($D$). The keys to the system are likely hiding in 
plain sight, waiting to be sampled.


{
  
    \bibliographystyle{plain} 
    \bibliography{references}

\begin{thebibliography}{10}

\bibitem{akiba2019optuna}
Takuya Akiba, Shotaro Sano, Toshihiko Yanase, Takeru Ohta, and Masanori Koyama.
\newblock Optuna: A next-generation hyperparameter optimization framework.
\newblock In {\em Proc. 25th ACM SIGKDD Int. Conf. Knowl. Discov. Data Min.}, pages 2623--2631, 2019.

\bibitem{Arthur2007}
David Arthur and Sergei Vassilvitskii.
\newblock {k}-means++: The advantages of careful seeding.
\newblock In {\em Proc. 18th Annu. ACM-SIAM Symp. Discrete Algorithms}, pages 1027--1035, 2007.

\bibitem{awad2021dehb}
Noor Awad, Neeratyoy Mallik, and Frank Hutter.
\newblock Dehb: Evolutionary hyperband for scalable, robust and efficient hyperparameter optimization.
\newblock In {\em Proc. Int. Joint Conf. Artif. Intell.}, pages 2147--2153, 2021.

\bibitem{Bergstra11}
James Bergstra, R{\'e}mi Bardenet, Yoshua Bengio, and Bal{\'a}zs K{\'e}gl.
\newblock Algorithms for hyper-parameter optimization.
\newblock In {\em Adv. Neural Inf. Process. Syst.}, pages 2546--2554, 2011.

\bibitem{bingham2019pyro}
Eli Bingham, Jonathan~P. Chen, Martin Jankowiak, Fritz Obermeyer, Neeraj Pradhan, et~al.
\newblock Pyro: deep universal probabilistic programming.
\newblock {\em J. Mach. Learn. Res.}, 20(1):973--978, 2019.

\bibitem{Brewka2011}
Gerhard Brewka, Thomas Eiter, and Miroslaw Truszczynski.
\newblock Answer set programming at a glance.
\newblock {\em Commun. ACM}, 54(12), 2011.

\bibitem{Bylander1991}
Tom Bylander, Dean Allemang, Michael~C. Tanner, and John~R. Josephson.
\newblock The computational complexity of abduction.
\newblock {\em Artif. Intell.}, 49, 1991.

\bibitem{carpenter2017stan}
Bob Carpenter, Andrew Gelman, Matthew~D Hoffman, Daniel Lee, Ben Goodrich, Michael Betancourt, Marcus Brubaker, Jiqiang Luo, Peter Li, and Allen Riddell.
\newblock Stan: A probabilistic programming language.
\newblock {\em Journal of statistical software}, 76(1), 2017.

\bibitem{Chen26Promise}
Pengzhou Chen and Tao Chen.
\newblock Promisetune: Unveiling causally promising and explainable tuning.
\newblock In {\em Proc. 48th Int. Conf. Softw. Eng.}, 2026.

\bibitem{cohen1995empirical}
Paul~R. Cohen.
\newblock {\em Empirical Methods for Artificial Intelligence}.
\newblock MIT Press, 1995.

\bibitem{ernst2016}
Neil~A. Ernst, Michael Popeck, Felix Bachmann, and Patrick Donohoe.
\newblock Creating software modernization roadmaps.
\newblock In {\em Proc. 13th Working IEEE/IFIP Conf. Softw. Archit.}, 2016.

\bibitem{Feather02}
Martin~S. Feather and Tim Menzies.
\newblock Converging on the optimal attainment of requirements.
\newblock In {\em Proc. IEEE Int. Requirements Eng. Conf.}, pages 263--270, 2002.

\bibitem{datalite}
Kishan~Kumar Ganguly and Tim Menzies.
\newblock How low can you go? the data-light se challenge, 2025.

\bibitem{Hamill09}
Margaret Hamill and Katerina Goseva-Popstojanova.
\newblock Common trends in software fault and failure data.
\newblock {\em IEEE Transactions on Software Engineering}, 35(4):484--496, July-Aug 2009.

\bibitem{Hindle12}
A.~Hindle, E.~T. Barr, Z.~Su, M.~Gabel, and P.~Devanbu.
\newblock On the naturalness of software.
\newblock In {\em Proc. Int. Conf. Softw. Eng. (ICSE)}, pages 837--847, 2012.

\bibitem{hutter2011sequential}
F.~Hutter, H.~H. Hoos, and K.~Leyton-Brown.
\newblock Sequential model-based optimization for general algorithm configuration.
\newblock In {\em Proc. Int. Conf. Learning and Intelligent Optimization (LION)}, pages 507--523, 2011.

\bibitem{Jureta2008}
I.~Jureta, J.~Mylopoulos, and S.~Faulkner.
\newblock Revisiting the core ontology and problem in requirements engineering.
\newblock In {\em Proc. IEEE Int. Requirements Eng. Conf. (RE)}, pages 71--80, 2008.

\bibitem{ref41}
M.~Li, T.~Chen, and X.~Yao.
\newblock How to evaluate solutions in pareto-based search-based software engineering: A critical review and methodological guidance.
\newblock {\em IEEE Trans. Softw. Eng.}, 48(5):1771--1799, 2020.

\bibitem{Lin15}
Z.~Lin and J.~Whitehead.
\newblock Why power laws? an explanation from fine-grained code changes.
\newblock In {\em Proc. Int. Conf. Mining Softw. Repos. (MSR)}, pages 68--75, 2015.

\bibitem{Long24}
D.~Long, S.~Drylie, J.~D. Ritschel, and C.~Koschnick.
\newblock An assessment of rules of thumb for software phase management.
\newblock {\em IEEE Trans. Softw. Eng.}, 50(2), 2024.

\bibitem{martinez2014fix}
Matias Martinez, Westley Weimer, and Martin Monperrus.
\newblock Do the fix ingredients already exist? an empirical inquiry into the redundancy assumptions of program repair approaches.
\newblock In {\em Companion Proceedings of the 36th international conference on software engineering}, pages 492--495, 2014.

\bibitem{Mathews17}
G.~Mathew, T.~Menzies, N.~A. Ernst, and J.~Klein.
\newblock Shorter: Reasoning about larger requirements models.
\newblock In {\em Proc. IEEE 25th Int. Requirements Eng. Conf. (RE)}, pages 154--163, 2017.

\bibitem{McGeoch2012}
C.~C. McGeoch.
\newblock {\em A Guide to Experimental Algorithmics}.
\newblock Cambridge Univ. Press, 2012.

\bibitem{MenziesKeys21}
T.~Menzies.
\newblock Shockingly simple: ``keys'' for better ai for se.
\newblock {\em IEEE Softw.}, 38(1):4--8, 2021.

\bibitem{Menzies2025MOOT}
T.~Menzies, T.~Chen, Y.~Ye, K.~K. Ganguly, A.~Rayegan, S.~Srinivasan, and A.~Lustosa.
\newblock Moot: A repository of many multi-objective optimization tasks.
\newblock arXiv:2511.16882, 2025.

\bibitem{michael1997uniformity}
C.~C. Michael and R.~C. Jones.
\newblock On the uniformity of error propagation in software.
\newblock In {\em Proc. 12th Annu. Comput. Assurance Conf. (COMPASS)}, pages 68--76, 1997.

\bibitem{miller1956magical}
George~A Miller.
\newblock The magical number seven, plus or minus two: Some limits on our capacity for processing information.
\newblock {\em Psychological review}, 63(2):81--97, 1956.

\bibitem{Ostrand04}
T.~J. Ostrand, E.~J. Weyuker, and R.~M. Bell.
\newblock Where the bugs are.
\newblock In {\em Proc. Int. Symp. Softw. Test. Anal. (ISSTA)}, 2004.

\bibitem{pelanek2008properties}
R.~Pel{\'a}nek.
\newblock Properties of state spaces and their applications.
\newblock {\em Int. J. Softw. Tools Technol. Transfer}, 10:443--454, 2008.

\bibitem{Peters15}
F.~Peters, T.~Menzies, and L.~Layman.
\newblock Lace2: Better privacy-preserving data sharing for cross-project defect prediction.
\newblock In {\em Proc. IEEE/ACM 37th Int. Conf. Softw. Eng. (ICSE)}, volume~1, pages 801--811, 2015.

\bibitem{Poole1994}
David Poole.
\newblock Who chooses the assumptions?
\newblock In {\em Abductive Reasoning}. Citeseer, 1994.

\bibitem{Simon1956}
H.~A. Simon.
\newblock Rational choice and the structure of the environment.
\newblock {\em Psychol. Rev.}, 63(2):129--138, 1956.

\bibitem{Selman03}
R.~Williams, C.~P. Gomes, and B.~Selman.
\newblock Backdoors to typical case complexity.
\newblock In {\em Proc. Int. Joint Conf. Artif. Intell. (IJCAI)}, pages 1173--1178, 2003.

\bibitem{wong1995reducing}
W.~Eric Wong and Aditya~P. Mathur.
\newblock Reducing the cost of mutation testing: An empirical study.
\newblock {\em J. Systems and Software}, 31:185--196, 1995.

\bibitem{Zave1997}
Pamela Zave and Michael Jackson.
\newblock Four dark corners of requirements engineering.
\newblock {\em ACM TOSEM}, 1997.

\bibitem{Zhang20BigFuzz}
Qian Zhang, Jiyuan Wang, Muhammad~Ali Gulzar, Rohan Padhye, and Miryung Kim.
\newblock Bigfuzz: Efficient fuzz testing for data analytics...
\newblock In {\em ASE '20}, pages 722--733, 2020.

\end{thebibliography}
}

 \end{document}